\documentclass[conference]{IEEEtran}
\IEEEoverridecommandlockouts
\usepackage{cite}
\usepackage{amsmath,amssymb,amsfonts}
\usepackage{algorithmicx}
\usepackage{algorithm}
\usepackage{algpseudocode}
\usepackage{graphicx}
\usepackage{textcomp}
\usepackage{xcolor}
\usepackage{placeins}
\usepackage{listings}
\usepackage{placeins}
\usepackage{url}
\usepackage{bytefield}
\def\BibTeX{{\rm B\kern-.05em{\sc i\kern-.025em b}\kern-.08em
    T\kern-.1667em\lower.7ex\hbox{E}\kern-.125emX}}
\definecolor{mGreen}{rgb}{0,0.6,0}
\definecolor{mGray}{rgb}{0.5,0.5,0.5}
\definecolor{mPurple}{rgb}{0.58,0,0.82}
\definecolor{backgroundColour}{rgb}{250,240,250}

\lstdefinestyle{CStyle}{
	backgroundcolor=\color{backgroundColour},   
	commentstyle=\color{mGreen},
	keywordstyle=\color{blue},
	numberstyle=\tiny\color{mGray},
	stringstyle=\color{mPurple},
	basicstyle=\footnotesize,
	breakatwhitespace=false,         
	breaklines=true,                 
	captionpos=b,                    
	keepspaces=true,                 
	numbers=left,                    
	numbersep=-3pt,                  
	showspaces=false,                
	showstringspaces=false,
	showtabs=false,                  
	tabsize=2,
	language=C
}

\usepackage{tikz-qtree}
\usetikzlibrary{shadows,trees}
\begin{document}

\title{RARES: \underline{R}untime \underline{A}ttack \underline{R}esilient \underline{E}mbedded \underline{S}ystem Design Using Verified Proof-of-Execution. \\}

\author{\IEEEauthorblockN{Avani Dave Nilanjan Banerjee Chintan Patel}

\IEEEauthorblockA{Department of Computer Science and Electrical Engineering, University of Maryland, Baltimore County}
}
{
}



\maketitle
\begin{abstract}
Modern society is getting accustomed to the Internet of Things (\textit{IoT}) and Cyber-Physical Systems (\textit{CPS}) for a variety of applications that involves security-critical user data and information transfers. In the lower end of the spectrum,  these devices are resource-constrained with no attack protection. They become a soft target for malicious code modification attacks that steals and misuses device data in malicious activities. The resilient system requires continuous detection, prevention, and/or recovery and correct code execution (including in degraded mode). By end large, existing security primitives (e.g., secure-boot, Remote Attestation \textit{RA}, Control Flow Attestation (\textit{CFA}) and Data Flow Attestation (\textit{DFA})) focuses on detection and prevention, leaving the proof of code execution and recovery unanswered. 
\par To this end, the proposed work presents lightweight \textit{RARES} - \underline{R}untime \underline{A}ttack \underline{R}esilient \underline{E}mbedded \underline{S}ystem design using verified Proof-of-Execution. It presents first custom hardware control register  (\texttt{Ctrl\_register}) based runtime memory modification attacks classification and detection technique. It further demonstrates the Proof Of Concept (\textit{POC}) implementation of use-case-specific attacks prevention and onboard recovery techniques. The prototype implementation on Artix~7 Field Programmable Gate Array (\textit{FPGA}) and state-of-the-art comparison demonstrates very low (2.3\%) resource overhead and efficacy of the proposed solution.   
\end{abstract}

\section{Introduction}
Industry 4.0 \cite{Ge:2013} has proliferated the use of connected small Internet of Things (IoT) and Cyber-Physical Systems (\textit{CPS}) in applications ranging from home security systems, smart controllers, actuators, sensor nodes, activity trackers, and alarm systems. Often these devices are used for security-critical user data and information transfers. A majority of them are resource-constrained \cite{openmsp:430, atmega:32}, with no onboard security support, which makes them vulnerable to code modification attacks. For example, the Electric Control Unit (ECU) of car measures various sensors (e.g., humidity, speed, temperature, speed) and performs different actuation tasks such as speed or heat controls.  If an attacker modifies the temperature sensor code to give a low reading, it can overheat the car or damage other parts. Here are few more examples \cite{furtak:2014, Stuxnet, jeep} of such attacks. \par
\begin{figure}[h]
	\begin{center}
		\includegraphics[width=3.5in]{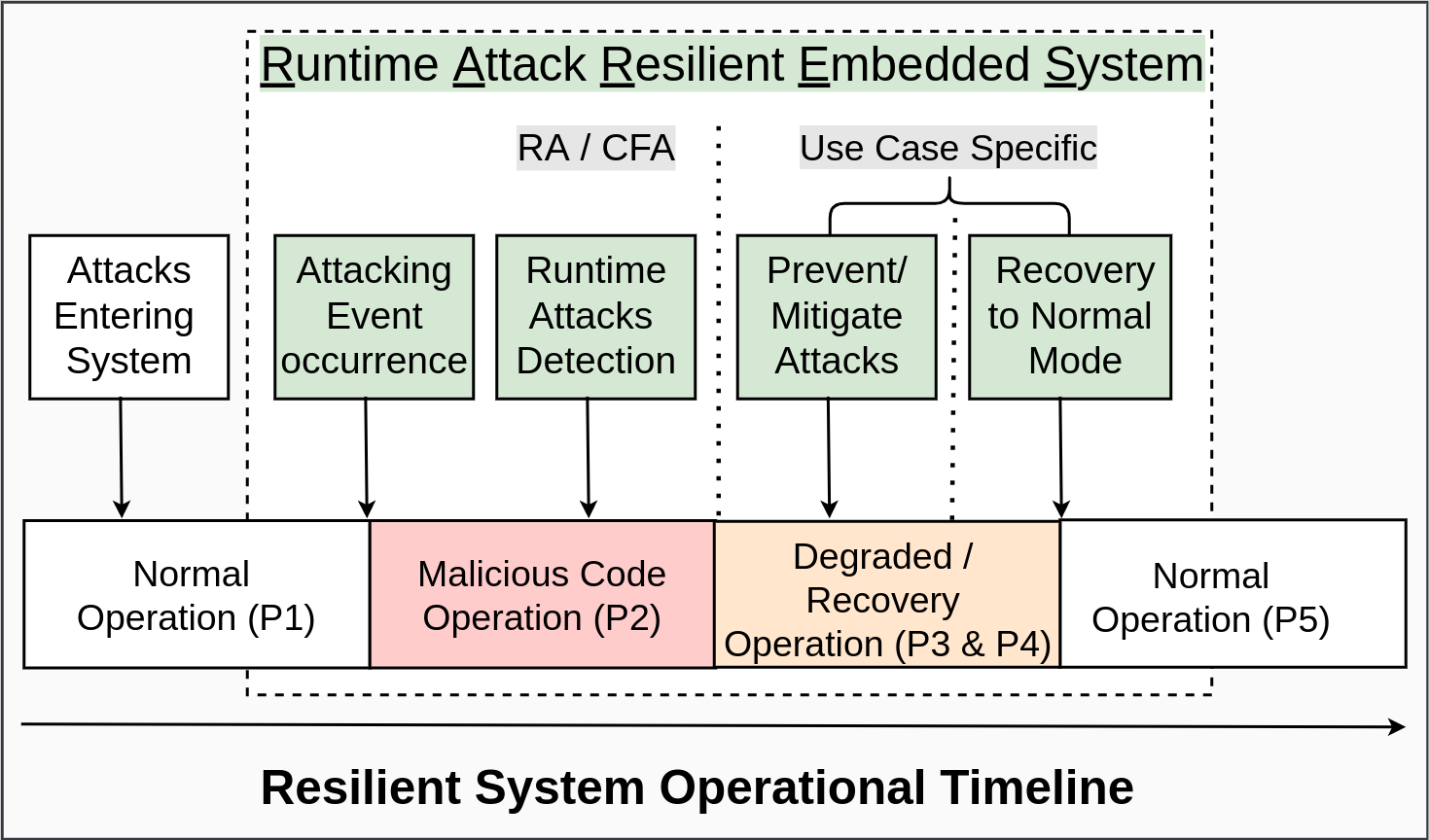}
	\end{center}
	\caption{ Depicts the resilient system operation flow timeline.  
		\label{fig:timeline}}
\end{figure}

 The resilience of a system is defined as its ability to detect (including boot-time and continuous runtime) the presence of attacks, prevent adversarial effects and keep the device operational (including in degraded mode) before it can reach a fail-safe or recovery state. Fig~\ref{fig:timeline} shows the resilient system operational timeline. The phases P1 and P5 depict the normal mode of operation.  Phase P2 covers attack occurrence and runtime detection. The phases-3 and 4 (P3 \& P4) represent the prevention and recovery operations.  
\par From Fig~\ref{fig:timeline}, the resilient system requires a secure-boot \cite{Haj:2019, NSA,Sanctum:2018, Wong:2018} like boot-time software integrity validation technique before the device enters in phase P1. Recent implementations of \textit{APEX} \cite{Apex:2020} presents lightweight continuous runtime attacks detection, prevention and verified proof of execution techniques (covering phase 2 and 3 from Fig~\ref{fig:timeline}). However, it resets the systems abruptly for attack prevention. Considering the wide applications spectrum of these devices (e.g., aircraft controllers, automotive Electronic Control Unit (\textit{ECU})), \label{intro} an abrupt system reset can result in adverse effects. These devices requires to operate (including degraded mode) until they can fail-safe or recover completely. Furthermore, they requires use-case-specific prevention and recovery techniques. \par To this end, this paper presents \textit{RARES}: a novel lightweight \underline{R}untime \underline{A}ttack \underline{R}esilient \underline{E}mbedded \underline{S}ystem design using verified proof of execution. \\
{\bf{Research Contributions:}} The design and implementation of \textit{RARES} presents the following research contributions:
\begin{itemize}
	\item{\bf{Runtime Attacks Classification \& Detection}:} It classifies runtime memory modification attacks into three categories and presents a novel lightweight 16 bit control register (\texttt{Ctrl\_register}) based detection technique.
	\item {\bf{Prevention Technique}:} It demonstrates two novel use-case-specific runtime attacks prevention techniques. It gives the control in the hands of system developers to design use-case-specific prevention and recovery techniques.
	\item {\bf{Secureboot and Recovery Technique}:} It presents the lightweight implementation of secure-boot and onboard recovery architecture for the resilient system. 
\end{itemize}

\section{Related work}
As shown in Fig~\ref{fig:timeline}, the resilient system design involves various phases of detection, prevention, and recovery. Unfortunately, \textit{RARES} was unable to find a single state-of-art implementation supporting all of these. Therefore, we have studied and analyzed the state-of-the-art works in three main categories: 1) detection, 2) prevention, and 3) recovery techniques. 
\subsubsection{Detection Techniques}
 Remote Attestation (\textit{RA}) is widely used client-server based security primitive that performs integrity verification of software state of the un-trusted prover device upon request from third party trusted verifier. Previous implementations of hardware-based (\cite{Tpm:2010},  \cite{Jiang:2017}, \cite{Sabt:2015},\cite{keystone}), software-based (\cite{Chakraborty:2019}, \cite{Goldman}) and hybrid (\cite{Wong:2018},\cite{Vrased:2019}) \textit{RA}s can detect runtime memory modification attacks periodically. Control Flow Attestation (\textit{CFA}) \cite{cflat:2016,Atrium:2017,lofat:2017, Lithex:2018} and Data Flow Attestation  (\textit{DFA})\cite{dflow:2006,dflowkernal:2016,dflow:2019} techniques are used for continuous runtime attacks detection. 
\subsubsection{Prevention Techniques}
 The resource isolation is well-known technique to prevent / limit the adverse effect of attacks. The hardware-based techniques uses Trusted Platform Module (\textit{TPM}) \cite{Tpm:2010}, Arm TrustZone \cite{Jiang:2017}, Trusted Execution Environment (\textit{TEE}) \cite{Sabt:2015}, or Physical Memory Protection (\textit{PMP}) \cite{Sracare:2020} to isolate the shared resources and limit the attacking surface. By end large, these are resource-heavy techniques and not suitable for targeted devices. Recent lightweight implementation of \textit{VRASED} \cite{Vrased2019} (formally verified remote attestation) uses custom hardware module to detect different security property based attacks. \textit{APEX} \cite{Apex:2020} extends \textit{VRASED} to provide verified Proof Of eXecution (\textit{POX}). They both resets the system to prevent the runtime attack. \textit{RARES} advocates the development of use-case-specific prevention or recovery techniques to avoid adverse effects from abrupt system reset. The detailed system design is discussed in subsection~\ref{design}.
\subsubsection{Recovery Techniques}
The affected device can be recovered by Over-The-Air (\textit{OTA}) or manually code re-flash. Recent implementation of \textit{Healed} \cite{Healed:2019} presents Merkle Hash Tree (\textit{MHT}) based technique, which requires at least one node in the network to be untampered, and its firmware is used to re-flash the corrupted node. \cite{Secerase:2010} keeps the software receiver-transmitter code in trusted \textit{ROM} for connecting the affected device to a recovery server for re-flash. Recent implementations of \textit{CARE} \cite{Care:2020} presents lightweight secure-boot with onboard recovery technique for the system where manual or over-the-air code reflash are not possible. \textit{SRACARE} \cite{Sracare:2020} extends \textit{CARE} by adding secure communication and \textit{RA} capabilities.
\par In summary, \textit{RA} can only detect periodic runtime attacks and it suffers from $CWE~367$-Time-of-Check-Time-of-Use (TOCTOU) \cite{toctou:2006} attacks. Both \textit{CFA} and \textit{DFA} bloats the system memory by storing runtime execution flow logs, which makes them unsuitable for targeted low-end devices. The lightweight runtime attacks detection technique presented by \textit{APEX} provides only one solution of resetting the system for preventing all different attacks. Furthermore, they do not cover the boot-time attacks detection or recovery techniques. \par Therefore, \textit{RARES} proposes the first implementation of the lightweight novel control register (\texttt{Ctrl\_register}) based runtime attacks detection, application-specific prevention techniques. Additionally, it presents the lightweight implementation of onboard recovery and secure-boot for resilient system design.  

\section{$RARES$ Overview}
This section covers the details about the targeted platform, \textit{RARES} based system architecture, design, and operation. 
\subsection{Targeted Platform} \label{design1}
 The low-end microcontrollers (e.g., Texas Instrument's MSP430 or Atmel AVR ATMega micro-controllers) are widely used in applications ranging from automotive ECU's, industrial control systems, actuators, aviation, sensors, smart \textit{IoT}, and Cyber-Physical System (\textit{CPS}). These devices have very low hardware foot print with only a few KB of address and data memories. They do not have sophisticated hardware or \textit{OS} support to detect and/or prevent the malware attacks. Therefore, \textit{RARES} was designed targeting the OpenMSP430 platform as well-maintained open cores implementation of OpenMSP430 \cite{openmsp:430} was readily available.  However, the proposed concept of custom control register (Ctrl\_register) based continuous runtime attacks classification and detection, use-case-specific prevention, and onboard recovery can be applied to other low-end devices such as Atmel AVR ATmega.  
\subsection{\textit{RARES} Design} \label{design}
\begin{figure}[h]
	\begin{center}
		\includegraphics[width=3.5in]{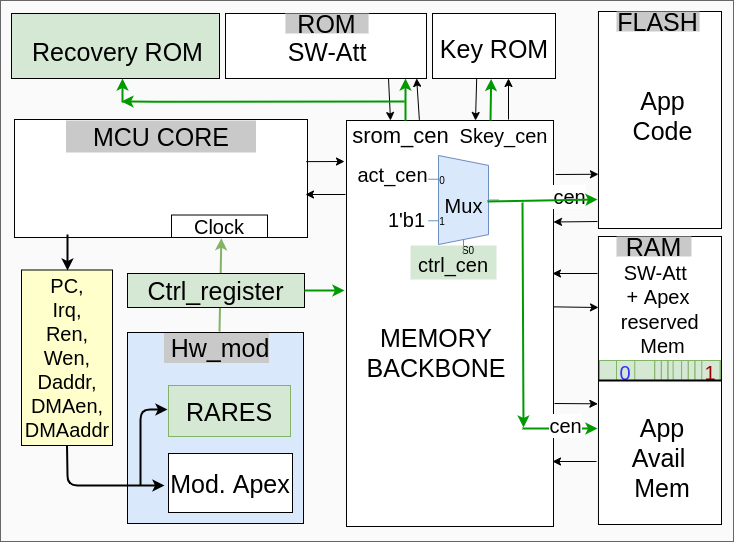}
	\end{center}
	\caption{Top-level design of lightweight runtime attacks resilient \textit{RARES} system. Highlighted are the key components of the proposed system.  
		\label{fig:soc}}
\end{figure}

Fig~\ref{fig:soc} shows the high-level design architecture of \textit{RARES}.
\FloatBarrier
\FloatBarrier
\textit{RARES} was designed on top of \textit{APEX}. It leverages underlying architecture to provide verified proof-of-execution (\textit{POX}). \textit{RARES} tapes out the seven control signals ($Pc$, $Irq$, $Ren$, $Wen$, $Daddr$, $DMAen$, $DMAaddr$) to its custom hardware module  (Hw\_mod). It has carefully designed and modified the internal Finite State Machines (FSMs) of both \textit{VRASED} and \textit{APEX} for detecting different categories of attacks in only one machine clock cycle (mclk), as discussed in subsection~\S\ref{attacks}. It stores different attack bits in 16-bit control register \texttt{Ctrl\_register}, which are discussed in detail in subsection~\S\ref{detect}. The \texttt{Ctrl\_register} does not have high-level write Application Program Interface (\textit{API}) access. The memory backbone acts as arbitration between the front end, \textit{DMA}, and execution-unit interfaces for any system memory (e.g., program, data, and peripheral) accesses, and it is used by \textit{RARES} for attack prevention. \textit{RARES} includes separate secure recovery \textit{ROM} for onboard recovery technique implementation as discussed in subsection~\ref{recov}.  
\subsection{\textit{RARES} Operation} \label{opn} Upon power-on the first-stage boot-loader (\textit{FSBL}) code (from \textit{ROM}) gets executed in reserved \textit{RAM} memory and performs the secure-boot verification on flash image. It re-flashes the corrupted flash memory using recovery image upon integrity failure detection, else the system operates normally.  \textit{RARES} satisfies all the security properties of \textit{APEX} and uses the formally verified software HMC\_SHA256 code (\textit{SW-Att} (HACL*) \cite{Hacl}) from \textit{ROM} for secure-boot and \textit{RA} functionality. After that, the test application code (App Code) from flash gets executed in a specific region of (App. Avail. Mem.) \textit{RAM} as shown in Fig~\ref{fig:soc}. (Due to the page limitations here, interested readers are requested to refer \cite{Apex:2020} for \textit{RA} and \textit{POX} operation). \textit{RARES} performs use-case-specific prevention and recovery operation as discussed in section~\ref{prev} upon runtime attack detection. 
\section{Runtime Attacks Classification} \label{attacks}
Based on the seven control signals ($Pc$, $Irq$, $Ren$, $Wen$, $Daddr$, $DMAen$, $DMAaddr$) input to the custom hardware module and security properties of \textit{APEX}, \textit{RARES} has classified memory modification attacks in three categories, namely: 1) \textit{CPU} access violation, 2) \textit{DMA} access violation, and 3) 
\FloatBarrier
 \begin{figure}
\begin{tikzpicture}
	[-,thick,%
	every node/.style={shape=rectangle,inner sep=3pt,draw,thick}]
	\footnotesize
	\node {Runtime Memory Modification Attacks Classification} [edge from parent fork down]
	[sibling distance=3cm]
	child {node {CPU Access Violation}
		[sibling distance=1cm]
		child {node {ROM}
		[sibling distance=1cm]
		child {node {R}}}
	    child {node {Stack}
	    [sibling distance=1cm]
	    child {node {R}}}
	    child {node {RAM}
    [sibling distance=1cm]
    child {node {R}}
child {node {W}}}}
	child {node {DMA Access Violation}
		[sibling distance=1cm]
		child {node {ROM}
			[sibling distance=1cm]
			child {node {R}}}
		child {node {Stack}
			[sibling distance=1cm]
			child {node {R}}}
		child {node {RAM}
			[sibling distance=1cm]
			child {node {R}}
			child {node {W}}}}
	child {node {Atomicity Violation}
		[sibling distance=1cm]
		child {node {RAM}}
		child {node {Stack}}
	};
\end{tikzpicture}
\caption{Runtime Memory Modification Attacks Classification}\label{class}
\end{figure}
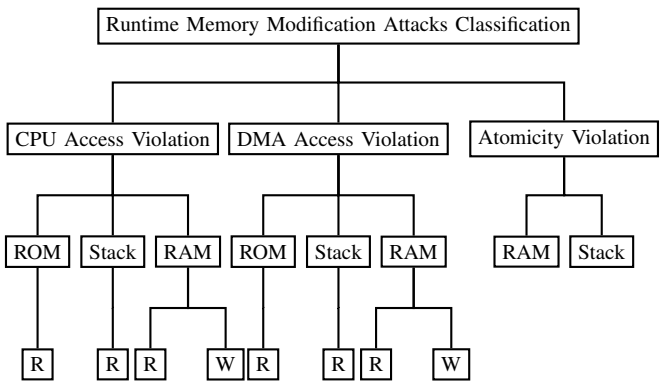
\FloatBarrier
Atomicity violation as shown in Fig~\ref{class}.
\subsubsection{CPU Access Violation} For the system shown in Fig~\ref{fig:soc}, while executing the program code from \textit{RAM} the \textit{CPU} can only read the data from reserved stack and \textit{ROM} (\textit{Sw-Att} code). However, it cannot access the device's secret key (\textit{K}) from the key \textit{ROM}. Similarly, The key (\textit{K}) is only accessed by the \textit{CPU} while it is executing the (\textit{Sw-Att}) code inside the reserved stack. All other \textit{ROM} and stack read accesses are detected as \textit{CPU} access violation by the hardware FSM in \texttt{Ctrl\_register}. Furthermore, any unauthorized \textit{RAM} access (both read and write) violation during \textit{Sw-Att} code execution are detected as \textit{CPU} related \textit{RAM} access violations. This sub-module focuses on ($Pc$, $Ren$, $Wen$, $Daddr$) control signals to detect any unauthorized memory read or write access request by the \textit{CPU}. The corresponding detection bits are updated in \texttt{Ctrl\_register} as discussed in subsection~\ref{detect}.
\subsubsection{DMA Access Violation} During the program code execution from \textit{RAM}, direct memory access (\textit{DMA}) read request from the reserved stack and \textit{ROM} (\textit{Sw-Att} code) are allowed. However, \textit{DMA} cannot access the device secret key (\textit{K}), while executing the program code from \textit{RAM}. Similarly, the \textit{DMA} can access the key (\textit{K}) only during \textit{Sw-Att} code execution inside the reserved stack. All other \textit{ROM} and stack related read accesses are identified as \textit{DMA} access violation by the hardware FSMs and detected in \texttt{Ctrl\_register}. Furthermore, unauthorized \textit{RAM} access (both read and write) violations while running  \textit{Sw-Att} code are detected under \textit{DMA}-related \textit{RAM} access violation. This sub-module focuses on ($Pc$, $Ren$, $Wen$, $DMAen$, $DMAaddr$) control signals to detect any unauthorized memory read or write access request using \textit{DMA}. The corresponding detection bits are updated in \texttt{Ctrl\_register} as discussed in subsection~\ref{detect}.
\subsubsection{Atomicity Violation} This category detects any interrupt trigger violation during the code execution inside \textit{RAM} and reserved stack (\textit{Sw-Att}). The atomicity violation usually results in interrupt service routine (IRQ) code execution, intermittent data and secure key (\textit{K}) leakage or loss. This sub-module detects mainly ($Irq$) IRQ during the code execution from the \textit{RAM} and reserved stack (\textit{Sw-Att}). The \textit{POC} atomicity violation prevention technique is discussed in subsection~\ref{chipen}.
\section{Detection Technique} \label{detect}
\par
\begin{figure}[h]
 		\begin{center}
 			\includegraphics[width=3.5in]{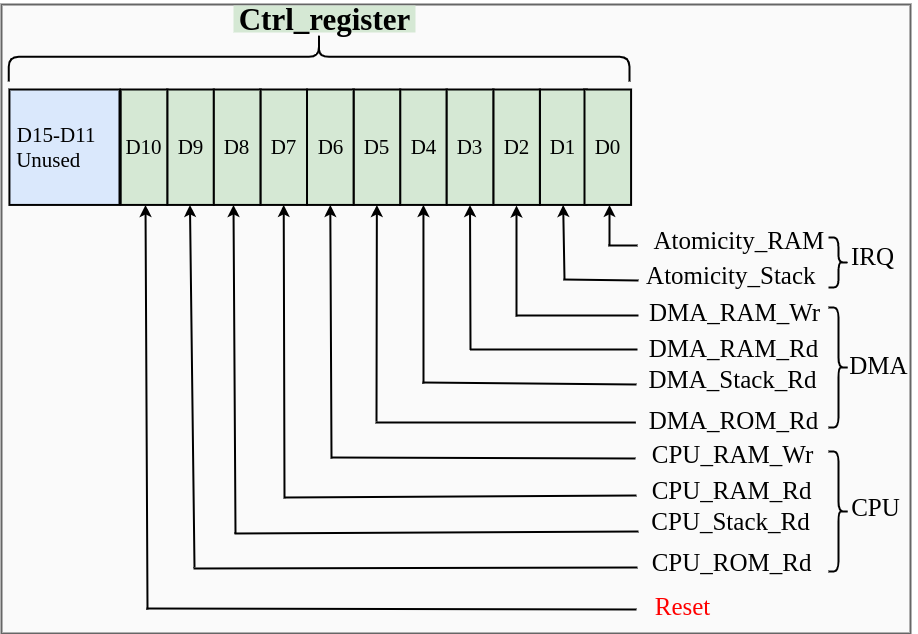}
 		\end{center}
 		\caption{Depicts 16 bit \texttt{Ctrl\_register} for each attacks detection by \textit{RARES}. Note that only 11-bits of the 16-bit \texttt{Ctrl\_register} are used currently, and remaining D11-D15 bits are left for future development.   
 			\label{fig:soc1}}
 	\end{figure}

 Based on attacks classification of section~\S\ref{attacks}, specific attack detection bits are updated in 16-bit \texttt{Ctrl\_register} as depicted in Fig~\ref{fig:soc1}. Note that, at current stage \textit{RARES} has classified and detected total ten different types of memory modification attacks and stored them in bit positions D0-D9.  
 The \texttt{Ctrl\_register} bit (D0) and (D1) detects atomicity violations during \textit{RAM} and stack code execution. The \textit{DMA} related \textit{RAM} write access violation is detected by flag bit (D2). The \textit{DMA} read access violations for \textit{RAM},stack, and \textit{ROM} are detected in bits (D3) (D4) and (D5), respectively. Similarly, \textit{CPU} related \textit{RAM} write access violation is detected by flag bit (D6). \textit{CPU} read access violations for \textit{RAM},stack, and \textit{ROM} are detected in bits (D7) (D8) and (D9), respectively. From this point the system developer can write use-case-specific runtime attack prevention or recovery technique.  
\section{Use-case Specific Prevention Techniques} \label{prev}
\textit{RARES} detects ten types of memory modification attacks in by setting corresponding flag bit high in \textit{Ctrl\_register}. It enables the system developer to implement multiple use-case-specific attacks prevention and recovery solutions instead of abrupt system reset like in \textit{APEX}. This section presents two use-case-specific runtime attack prevention techniques, and onboard recovery to demonstrate the efficacy of the proposed solution. \\
The test application performs \textit{RA} feature for the integrity verification of the flash memory region. It uses hmac\_sha256 (\textit{Sw-Att}) for the digest computation. The key {K} is only accessed by the system during \textit{Sw-Att} code execution. \textit{RARES} has implemented two different attacks and their prevention techniques as follows.
\subsection{ROM Key (K) Read Attack Prevention} \label{mod1} 
\textit{RARES} has implemented key (\textit{ROM}) read attack, by making \textit{CPU} read key (K) location while executing code from \textit{RAM} memory. It gets detected by bit D9 in  Ctrl\_register (CPU\_ROM\_Rd). \textit{RARES} has implemented one hardware based and one software based prevention techniques.\\ {\bf{1. Software based Prevention}}: \textit{RARES} has identified that, the underlying OpenMSP430 micro-controller has six different 
\FloatBarrier
\vspace{-2em}
\begin{figure}
	\begin{lstlisting}[style=CStyle]
		void prev() 
		{
			__asm__ volatile("bis  #240, r2" "\n\t");  
		}
	\end{lstlisting}
	\caption{Depicts software-based \textit{RARES} based mode switching technique for attack prevention}
	\label{fig:mod}
\end{figure}
\FloatBarrier
Low Power Modes (e.g., $LPM0$, $LPM1$, $LMP2$, $LPM3$, $LPM4$, $LPM5$) for mainly power conservation. The system can be switched to operate in any of the available mode based on the value in r2 register. \textit{RARES} has leveraged this mode switching capabilities of the targeted devices to prevent runtime attacks. It has implemented and validated software-based mode switching upon ROM access violation detected by bit D9. \textit{RARES} was switched to \textit{LPM0} to prevent the read \textit{ROM} attack. Fig~\ref{fig:mod} shows the code snippet of software-based mode-switching.  \\
{\bf{2. Hardware based Prevention}}:
 \textit{RARES} detects the ROM access violation by setting bit D9 high in  \textit{Ctrl\_register}. \textit{RARES} has identified that the underlying OpenMSP430 microcontrollers has a hardware pin called \textit{CPUOFF}, which makes the \textit{CPU} goes into the idle state (not off) while keeping mclk (peripherals) and \textit{DMA} ON. For the \textit{POC} of runtime attack prevention, \textit{RARES} has ORed (bitwise logic OR operation) the D9 bit with the  \textit{CPUOFF} bit $ON$ selection logic in hardware. 
 \par Another use-case for this solution could be, consider a sample application in car \textit{ECU}, where the device is continuously reading data from attached sensor node using \textit{DMA} and \textit{CPU} is performing arithmetic computation on different data. In this case, the runtime \textit{CPU} related \textit{ROM} access violation is detected by \textit{Ctrl\_register}. The proposed technique becomes very useful as it keeps the connected peripherals and \textit{DMA} ON, while keeping \textit{CPU} in the idle state.   \\
\subsection{RAM write Access Prevention} \label{recov}
\par
\begin{figure}[h]
	\begin{center}
		\includegraphics[width=3.5in]{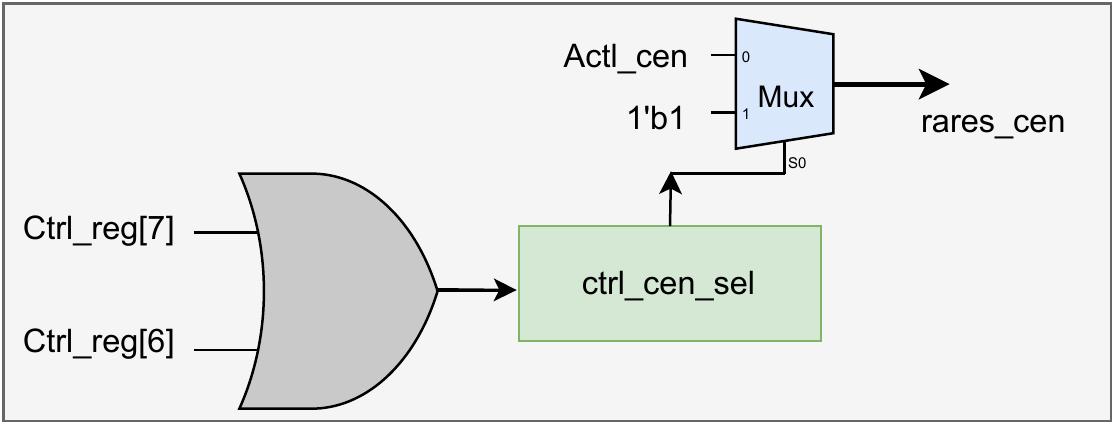}
	\end{center}
	\caption{Depicts unauthorized memory read/write attack prevention technique 
		\label{fig:soc3}}
\end{figure}

 To prevent runtime \textit{RAM} access (read/ write) violations, Fig~\ref{fig:soc3} demonstrates the first implementation of hardware-based prevention. 
\label{chipen} 
This solution was only possible because, \textit{RARES} was able to identified that underlying OpenMSP430 has individual (active low) chip enable signal for each memory modules (e.g., \texttt{srom\_cen}, \texttt{skey\_cen}, \texttt{pmem\_cen}, \texttt{dmem\_cen}) to trigger the data transfer, which are controlled by the memory backbone. \textit{RARES} uses the attack detection flag bits as chip enable selection switch. \textit{RARES} performs bitwise OR operation on corresponding memory access violations bits to generate the \texttt{ctrl\_cen\_sel} signal. \textit{RARES} makes the chip enable (chip\_en) signal high upon attack detection by \texttt{ctrl\_cen\_sel}. By doing this, \textit{RARES} inserts a wait states in current memory read/write instruction to pause the operation 
\FloatBarrier
\begin{table*}[h!]
	\centering 
	\caption{State-of-the-art (Qualitative) Comparison of Lightweight Attack Resilient Systems }
	\scalebox{1}{
		\begin{tabular} {@{}lcccccccccc@{}} \hline
			\multicolumn{1}{l|}{Parameters}        & \multicolumn{1}{c|}{{\bf{\textit{RARES}}}}&
			\multicolumn{1}{c|}{Ref.\cite{Apex:2020}} & \multicolumn{1}{c|}{Ref.\cite{Vrased:2019}} & \multicolumn{1}{c|}{Ref.\cite{Haj:2019}} & \multicolumn{1}{c|}{ref.\cite{Lithex:2018}} & \multicolumn{1}{c|}{Ref.\cite{lofat:2017}} & \multicolumn{1}{c}{Ref.\cite{Atrium:2017}} & \multicolumn{1}{c|}{Ref.\cite{Sracare:2020}} & \multicolumn{1}{c}{Ref.\cite{Healed:2019}} \\  \hline
			\multicolumn{1}{l|}{Design Type}         & \multicolumn{1}{c|}{Hybrid} &\multicolumn{1}{c|}{Hybrid} & \multicolumn{1}{c|}{Hybrid} & \multicolumn{1}{c|}{HW} & \multicolumn{1}{c|}{Hybrid} & \multicolumn{1}{c|}{Hybrid}   & \multicolumn{1}{c|}{Hybrid}   & \multicolumn{1}{c|}{Hybrid}   & \multicolumn{1}{c}{SW}  \\ 
			\multicolumn{1}{l|}{Secure Communication}   & \multicolumn{1}{c|}{yes}  & \multicolumn{1}{c|}{yes}  &\multicolumn{1}{c|}{yes} & \multicolumn{1}{c|}{no} &  \multicolumn{1}{c|}{yes} & \multicolumn{1}{c|}{yes}  & \multicolumn{1}{c|}{yes}  & \multicolumn{1}{c|}{yes}   & \multicolumn{1}{c}{no}  \\
			\multicolumn{1}{l|}{Lightweight}      & \multicolumn{1}{c|}{yes} &\multicolumn{1}{c|}{yes} & \multicolumn{1}{c|}{yes} & \multicolumn{1}{c|}{no} & \multicolumn{1}{c|}{yes} & \multicolumn{1}{c|}{yes}  & \multicolumn{1}{c|}{no} & \multicolumn{1}{c|}{yes}    & \multicolumn{1}{c}{no}  \\ 
			\multicolumn{1}{l|}{Secure boot}         & \multicolumn{1}{c|}{yes}  & \multicolumn{1}{c|}{no}  & \multicolumn{1}{c|}{no} & \multicolumn{1}{c|}{yes} &  \multicolumn{1}{c|}{no} & \multicolumn{1}{c|}{no} & \multicolumn{1}{c|}{no}  & \multicolumn{1}{c|}{yes}  & \multicolumn{1}{c}{no} \\
			\multicolumn{1}{l|}{Remote Attestation (periodic \textit{RA})}    & \multicolumn{1}{c|}{yes}      & \multicolumn{1}{c|}{yes}  & \multicolumn{1}{c|}{yes} &\multicolumn{1}{c|}{no} & \multicolumn{1}{c|}{no} & \multicolumn{1}{c|}{no} & \multicolumn{1}{c|}{no} & \multicolumn{1}{c|}{yes} & \multicolumn{1}{c}{yes}  \\
			\multicolumn{1}{l|}{Runtime Attacks Detection}         & \multicolumn{1}{c|}{yes}  & \multicolumn{1}{c|}{yes}  & \multicolumn{1}{c|}{yes} & \multicolumn{1}{c|}{no}  & \multicolumn{1}{c|}{yes} &  \multicolumn{1}{c|}{yes} & \multicolumn{1}{c|}{yes} & \multicolumn{1}{c|}{no}  & \multicolumn{1}{c}{no}   \\
			\multicolumn{1}{l|}{System Reset for Attacks Prevention}         & \multicolumn{1}{c|}{no}  & \multicolumn{1}{c|}{yes}  & \multicolumn{1}{c|}{yes} & \multicolumn{1}{c|}{yes} &  \multicolumn{1}{c|}{no} & \multicolumn{1}{c|}{no} & \multicolumn{1}{c|}{no}  & \multicolumn{1}{c|}{no} & \multicolumn{1}{c}{no}  \\ 
			\multicolumn{1}{l|}{Memory Mod. Attacks Prevention}         & \multicolumn{1}{c|}{yes}  & \multicolumn{1}{c|}{no}  & \multicolumn{1}{c|}{no} & \multicolumn{1}{c|}{no} &  \multicolumn{1}{c|}{no} & \multicolumn{1}{c|}{no} & \multicolumn{1}{c|}{no}  & \multicolumn{1}{c|}{no} & \multicolumn{1}{c}{no}  \\ 
			\multicolumn{1}{l|}{Recovery Techniques}   & \multicolumn{1}{c|}{yes}  & \multicolumn{1}{c|}{no}  &\multicolumn{1}{c|}{no}  & \multicolumn{1}{c|}{no} & \multicolumn{1}{c|}{no} &  \multicolumn{1}{c|}{no} & \multicolumn{1}{c|}{no} & \multicolumn{1}{c|}{yes} &
			\multicolumn{1}{c}{yes} \\ \hline
	\end{tabular}}\label{sa}
\end{table*}
\FloatBarrier
  by hardware (this covers phase P3 in Fig~\ref{fig:timeline}), while keeping other operations ON. Inserting wait state pauses the unauthorized code execution. However, to bring the device back to normal operation, \textit{RARES} triggers a subroutine call to performs the code reflash using un-tempered (golden) recovery (\textit{ROM}) image.  For the atomicity violations \textit{RARES} triggers system reset by enabling D10 reset signal.
\par Note that, the above prevention and recovery techniques are implemented for \textit{POC} only. The goal of \textit{RARES} design is to demonstrate runtime detection using the 16-bit \textit{Ctrl\_register}, and to give system developers an opportunity to design their use-case-specific prevention and recovery solution.\\


\section{Evaluation}
This section performs qualitative and quantitative evaluation of \textit{RARES} based resilient system. The subsection~\S\ref{reu} covers the resource utilization (and overheads) for quantitative analysis and subsection~\S\ref{stcom} performs the state-of-the-art comparison for qualitative analysis.
\subsection{Resource Utilization - Quantitative Comparison}\label{reu}
\textit{RARES} was implemented on top of \textit{APEX} \cite{Apex:2020} and complete verilog Resistor Transistor Logic (\textit{RTL}) was synthesized on Artix-7 Field Programmable Gate Array (\textit{FPGA}) board using Xilinx Vivado~2018. As shown in Fig~\ref{fig:soc}, the new control register (Ctrl\_register) was added into \textit{APEX}'s \textit{METADATA} with only read access from the software. Therefore, \textit{RARES} increases the reserved memory of \textit{APEX} by two bytes to store 16-bit \texttt{Ctrl\_register}. 
\begin{table}[h!]
	\caption{Resource Utilization- Quantitative Comparison}\label{ru}
	\scalebox{.85}{
		\begin{tabular}{@{}lcccc@{}} 
			\hline
			\multicolumn{1}{l|}{Architecture}        & \multicolumn{1}{c}{Hardware} &	\multicolumn{1}{c|}{Resources} &  \multicolumn{1}{c|}{Reserved Mem.} & \multicolumn{1}{c}{Verified} \\
			\multicolumn{1}{l|}{Details}        & \multicolumn{1}{c|}{Reg.} &	\multicolumn{1}{c|}{LUT } &  \multicolumn{1}{c|}{RAM (bytes)} & \multicolumn{1}{c}{\# LTL } \\ \hline 
			\multicolumn{1}{l|}{OpenMSP430 \cite{openmsp:430}}    & \multicolumn{1}{c|}{691}  & \multicolumn{1}{c|}{1904} & \multicolumn{1}{c|}{0}  &\multicolumn{1}{c}{-}   \\
			\multicolumn{1}{l|}{$VRASED$ \cite{Vrased:2019}}      & \multicolumn{1}{c|}{729}  & \multicolumn{1}{c|}{1980} & \multicolumn{1}{c|}{2332} & \multicolumn{1}{c}{10}   \\
			\multicolumn{1}{l|}{$APEX$ \cite{Apex:2020}}         & \multicolumn{1}{c|}{755}  & \multicolumn{1}{c|}{2290} & \multicolumn{1}{c|}{2341} &  \multicolumn{1}{c}{20} \\
			\multicolumn{1}{l|}{\textit{$RARES-A$}}  & \multicolumn{1}{c|}{773}  & \multicolumn{1}{c|}{2330} & \multicolumn{1}{c|}{2343} &   \multicolumn{1}{c}{20}  \\
			\multicolumn{1}{l|}{$RARES-B$}  & \multicolumn{1}{c|}{830}  & \multicolumn{1}{c|}{2572} & \multicolumn{1}{c|}{2343} &   \multicolumn{1}{c}{20}  \\  
			\hline
	\end{tabular}}
\end{table}
Table~\ref{ru} shows the hardware and memory resource utilization for a \textit{RARES} based system and compares it with different state-of-the-art implementations. The baseline Openmsp430 has the lowest hardware resources and requires no reserved memory. \textit{VRASED} uses approximately 2~KB of reserved stack memory for computing the \textit{RA} (using SW-Att code) and storing results. \textit{APEX} adds nine bytes to store the verified proof of execution. \textit{RARES-A} extends it further by 2 bytes for storing 16-bit \texttt{Ctrl\_register} at runtime. \par This work has calculated hardware resource footprint for two implementations, 1) \textit{RARES-A} with 16-bit \texttt{Ctrl\_register} and 2) \textit{RARES-B} which includes the additional onboard recovery \textit{ROM}. \textit{RARES-A} requires 2.3\% more hardware registers and approximately 1.7\% more \textit{LUT} than \textit{APEX}. \textit{RARES-B} adds the recovery memory (as shown in Fig~\ref{fig:soc})  and it requires 7.37\% more hardware registers and approximately 10.3\% more \textit{LUT} than \textit{RARES-A} (for 16KB \textit{ROM}). \textit{RARES} maintains all twenty formal \textit{LTL} specification verification of \textit{APEX}. 
\subsection{State-of-the-art Qualitative Comparison} \label{stcom}
\textit{RARES} was compared with state-of-the-art secure-boot, remote attestation, control flow attestation, and recovery-based resilience systems for qualitative analysis as shown in Table~\ref{sa}. \textit{RA}s provide periodic runtime software state verification. 
\textit{CFA} and \textit{DFA} (\cite{Lithex:2018,lofat:2017,Atrium:2017}) provides continuous runtime attacks detection. However, they are resource-heavy and bloats the system memory by logs storing.  The lightweight implementations of \textit{APEX} \cite{Apex:2020}, \textit{VRASED} \cite{Vrased:2019}, and \cite{Haj:2019} resets the systems to prevent the attacks. Only \textit{RARES} based system offers lightweight runtime attacks detection, use-case-specific prevention, secure-boot and onboard recovery techniques without an abrupt system reset. 

\section{Discussion}
\textit{RARES} is the first implementation of runtime memory modification attacks detection using \textit{Ctrl\_register}. It opens a broad possibilities of use-case-specific attack prevention or recovery-based system design. Since \textit{RARES} is designed on top of \textit{APEX}, all the security properties and formal verification proofs are maintained, along with the adversarial model and limitations. The proposed solution can be generalized and applied to any lightweight micro-controllers by attaching the custom hardware module to them. While porting to the new platform will require the system developer to identify and implement the (new) platform specific suitable prevention and recovery techniques. The solution can be used with different test applications as well by storing the correct recovery image in \textit{ROM}. \textit{RARES} uses onboard recovery to cover broad application areas where over-the-air or manual recovery is not possible. The system designer can store only the critical section of flash code instead of the full image to reduce the resource utilization in \textit{RARES-B} implementation.
\section{Conclusion}
The lightweight attack resilient system design requires runtime memory modification attacks detection, prevention, and/or recovery techniques. \textit{RARES} demonstrates the first implementation and efficacy of a novel lightweight control register (Ctrl\_register) based continuous runtime attacks detection technique. This approach enables the system developers to design use-case-based prevention techniques. It further showcases two runtime memory modification attack prevention and onboard recovery techniques. It requires very little resource overhead when compared with the state-of-the-art techniques. 

\bibliographystyle{IEEEtran}

\bibliography{references}

\begin{thebibliography}{10}
\providecommand{\url}[1]{#1}
\csname url@samestyle\endcsname
\providecommand{\newblock}{\relax}
\providecommand{\bibinfo}[2]{#2}
\providecommand{\BIBentrySTDinterwordspacing}{\spaceskip=0pt\relax}
\providecommand{\BIBentryALTinterwordstretchfactor}{4}
\providecommand{\BIBentryALTinterwordspacing}{\spaceskip=\fontdimen2\font plus
\BIBentryALTinterwordstretchfactor\fontdimen3\font minus
  \fontdimen4\font\relax}
\providecommand{\BIBforeignlanguage}[2]{{%
\expandafter\ifx\csname l@#1\endcsname\relax
\typeout{** WARNING: IEEEtran.bst: No hyphenation pattern has been}%
\typeout{** loaded for the language `#1'. Using the pattern for}%
\typeout{** the default language instead.}%
\else
\language=\csname l@#1\endcsname
\fi
#2}}
\providecommand{\BIBdecl}{\relax}
\BIBdecl

\bibitem{Ge:2013}
\BIBentryALTinterwordspacing
H.~Kagermann, W.~Wahlster, and J.~Helbig, ``Recommendations for implementing
  the strategic initiative industrie 4.0 -- securing the future of german
  manufacturing industry,'' acatech -- National Academy of Science and
  Engineering, M\"{u}nchen, Final Report of the Industrie 4.0 Working Group,
  2013. [Online]. Available:
  \url{\url{http://forschungsunion.de/pdf/industrie_4_0_final_report.pdf}}
\BIBentrySTDinterwordspacing

\bibitem{openmsp:430}
O.~Girard, ``Openmsp430,'' \url{https://opencores.org/projects/openmsp430},
  2009.

\bibitem{atmega:32}
``8/16-bit atmel xmega b3 microcontroller,''
  \url{http://ww1.microchip.com/downloads/en/DeviceDoc/Atmel-8074-8-and-16-bit-AVR-Microcontroller-XMEGA-B-ATxmega64B3-ATxmega128B3_Datasheet.pdf},
  2014.

\bibitem{furtak:2014}
A.~Furtak, Y.~Bulygin, O.~Bazhaniuk, J.~Loucaides, A.~Matrosov, and M.~Gorobet,
  ``Bios and secure boot attacks uncovered,'' 2014.

\bibitem{Stuxnet}
J.~Vijayan, ``Stuxnet renews power grid security concerns,''
  \url{https://www.computerworld.com/article/2519574/stuxnet-renews-power-grid-security-concerns.html},
  June 2010.

\bibitem{jeep}
D.~Schneider, ``Jeep hacking 101,''
  http://spectrum.ieee.org/cars-that-think/transportation/systems/jeep-hacking-101.html,
  2015.

\bibitem{Haj:2019}
J.~{Haj-Yahya}, M.~M. {Wong}, V.~{Pudi}, S.~{Bhasin}, and A.~{Chattopadhyay},
  ``Lightweight secure-boot architecture for risc-v system-on-chip,'' in
  \emph{20th International Symposium on Quality Electronic Design (ISQED)},
  March 2019, pp. 216--223.

\bibitem{NSA}
{NSA Cyber Report}, ``{UEFI DEFENSIVE PRACTICES GUIDANCE},''
  \url{https://www.nsa.gov/Portals/70/documents/what-we-do/cybersecurity/professional-resources/ctr-uefi-defensive-practices-guidance.pdf},
  2017.

\bibitem{Sanctum:2018}
I.~{Lebedev}, K.~{Hogan}, and S.~{Devadas}, ``Invited paper: Secure boot and
  remote attestation in the sanctum processor,'' in \emph{2018 IEEE 31st
  Computer Security Foundations Symposium (CSF)}, July 2018, pp. 46--60.

\bibitem{Wong:2018}
M.~M. Wong, J.~Haj-Yahya, and A.~Chattopadhyay, ``Smarts: secure memory
  assurance of risc-v trusted soc,'' 06 2018, pp. 1--8.

\bibitem{Apex:2020}
\BIBentryALTinterwordspacing
I.~D.~O. Nunes, K.~Eldefrawy, N.~Rattanavipanon, and G.~Tsudik, ``{APEX}: A
  verified architecture for proofs of execution on remote devices under full
  software compromise,'' in \emph{29th {USENIX} Security Symposium ({USENIX}
  Security 20)}.\hskip 1em plus 0.5em minus 0.4em\relax {USENIX} Association,
  Aug. 2020, pp. 771--788. [Online]. Available:
  \url{https://www.usenix.org/conference/usenixsecurity20/presentation/nunes}
\BIBentrySTDinterwordspacing

\bibitem{Tpm:2010}
``{Trusted Platform Module - Trusted Computing Group},''
  \url{https://trustedcomputinggroup.org/work-groups/trusted-platform-module/},
  2010.

\bibitem{Jiang:2017}
H.~Jiang, R.~Chang, L.~Ren, and W.~Dong, ``Implementing a arm-based secure boot
  scheme for the isolated execution environment,'' in \emph{2017 13th
  International Conference on Computational Intelligence and Security (CIS)},
  Dec 2017, pp. 336--340.

\bibitem{Sabt:2015}
M.~{Sabt}, M.~{Achemlal}, and A.~{Bouabdallah}, ``Trusted execution
  environment: What it is, and what it is not,'' in \emph{2015 IEEE
  Trustcom/BigDataSE/ISPA}, vol.~1, Aug 2015, pp. 57--64.

\bibitem{keystone}
D.~Lee, D.~Kohlbrenner, S.~Shinde, D.~X. Song, and K.~Asanovic, ``Keystone: A
  framework for architecting tees,'' \emph{ArXiv}, vol. abs/1907.10119, 2019.

\bibitem{Chakraborty:2019}
\BIBentryALTinterwordspacing
D.~Chakraborty, L.~Hanzlik, and S.~Bugiel, ``simtpm: User-centric {TPM} for
  mobile devices,'' in \emph{28th {USENIX} Security Symposium ({USENIX}
  Security 19)}.\hskip 1em plus 0.5em minus 0.4em\relax Santa Clara, CA:
  {USENIX} Association, Aug. 2019, pp. 533--550. [Online]. Available:
  \url{https://www.usenix.org/conference/usenixsecurity19/presentation/chakraborty}
\BIBentrySTDinterwordspacing

\bibitem{Goldman}
{Goldman, Ken}, ``{IBM-ACS},''
  \url{https://sourceforge.net/p/ibmtpm20acs/activity/?page=0{\&}limit=100{\#}5cc3737aee24ca5b73320e9c},
  2017.

\bibitem{Vrased:2019}
\BIBentryALTinterwordspacing
I.~D.~O. Nunes, K.~Eldefrawy, N.~Rattanavipanon, M.~Steiner, and G.~Tsudik,
  ``{VRASED}: A verified hardware/software co-design for remote attestation,''
  in \emph{28th {USENIX} Security Symposium ({USENIX} Security 19)}.\hskip 1em
  plus 0.5em minus 0.4em\relax Santa Clara, CA: {USENIX} Association, Aug.
  2019, pp. 1429--1446. [Online]. Available:
  \url{https://www.usenix.org/conference/usenixsecurity19/presentation/de-oliveira-nunes}
\BIBentrySTDinterwordspacing

\bibitem{cflat:2016}
\BIBentryALTinterwordspacing
T.~Abera, N.~Asokan, L.~Davi, J.-E. Ekberg, T.~Nyman, A.~Paverd, A.-R. Sadeghi,
  and G.~Tsudik, ``C-flat: Control-flow attestation for embedded systems
  software,'' in \emph{Proceedings of the 2016 ACM SIGSAC Conference on
  Computer and Communications Security}, ser. CCS '16.\hskip 1em plus 0.5em
  minus 0.4em\relax New York, NY, USA: Association for Computing Machinery,
  2016, p. 743–754. [Online]. Available:
  \url{https://doi.org/10.1145/2976749.2978358}
\BIBentrySTDinterwordspacing

\bibitem{Atrium:2017}
S.~{Zeitouni}, G.~{Dessouky}, O.~{Arias}, D.~{Sullivan}, A.~{Ibrahim},
  Y.~{Jin}, and A.~{Sadeghi}, ``Atrium: Runtime attestation resilient under
  memory attacks,'' in \emph{2017 IEEE/ACM International Conference on
  Computer-Aided Design (ICCAD)}, 2017, pp. 384--391.

\bibitem{lofat:2017}
\BIBentryALTinterwordspacing
G.~Dessouky, S.~Zeitouni, T.~Nyman, A.~Paverd, L.~Davi, P.~Koeberl, N.~Asokan,
  and A.-R. Sadeghi, ``Lo-fat: Low-overhead control flow attestation in
  hardware,'' in \emph{Proceedings of the 54th Annual Design Automation
  Conference 2017}, ser. DAC '17.\hskip 1em plus 0.5em minus 0.4em\relax New
  York, NY, USA: Association for Computing Machinery, 2017. [Online].
  Available: \url{https://doi.org/10.1145/3061639.3062276}
\BIBentrySTDinterwordspacing

\bibitem{Lithex:2018}
G.~{Dessouky}, T.~{Abera}, A.~{Ibrahim}, and A.~{Sadeghi}, ``Litehax:
  Lightweight hardware-assisted attestation of program execution,'' in
  \emph{2018 IEEE/ACM International Conference on Computer-Aided Design
  (ICCAD)}, 2018, pp. 1--8.

\bibitem{dflow:2006}
M.~Castro, M.~Costa, and T.~Harris, ``Securing software by enforcing data-flow
  integrity,'' in \emph{Proceedings of the 7th Symposium on Operating Systems
  Design and Implementation}, ser. OSDI '06.\hskip 1em plus 0.5em minus
  0.4em\relax USA: USENIX Association, 2006, p. 147–160.

\bibitem{dflowkernal:2016}
C.~Song, B.~Lee, K.~Lu, W.~R. Harris, T.~Kim, and W.~Lee, ``Enforcing kernel
  security invariants with data flow integrity.'' in \emph{NDSS 2016}, 2016.

\bibitem{dflow:2019}
A.~R. RACHALA, ``Evaluation of hardware-based data flow integrity,''
  https://core.ac.uk/download/pdf/266596061.pdf, 2019.

\bibitem{Sracare:2020}
A.~{Dave}, N.~{Banerjee}, and C.~{Patel}, ``Sracare: Secure remote attestation
  with code authentication and resilience engine,'' in \emph{2020 IEEE
  International Conference on Embedded Software and Systems (ICESS)}, 2020, pp.
  1--8.

\bibitem{Vrased2019}
\BIBentryALTinterwordspacing
I.~D.~O. Nunes, K.~Eldefrawy, N.~Rattanavipanon, M.~Steiner, and G.~Tsudik,
  ``{VRASED}: A verified hardware/software co-design for remote attestation,''
  in \emph{28th {USENIX} Security Symposium ({USENIX} Security 19)}.\hskip 1em
  plus 0.5em minus 0.4em\relax Santa Clara, CA: {USENIX} Association, Aug.
  2019, pp. 1429--1446. [Online]. Available:
  \url{https://www.usenix.org/conference/usenixsecurity19/presentation/de-oliveira-nunes}
\BIBentrySTDinterwordspacing

\bibitem{Healed:2019}
A.~Ibrahim, A.-R. Sadeghi, and G.~Tsudik, ``Healed: Healing and attestation for
  low-end embedded devices,'' in \emph{Financial Cryptography}, 2019.

\bibitem{Secerase:2010}
D.~Perito and G.~Tsudik, ``Secure code update for embedded devices via proofs
  of secure erasure,'' in \emph{ESORICS}, 2010.

\bibitem{Care:2020}
A.~{Dave}, N.~{Banerjee}, and C.~{Patel}, ``Care: Lightweight attack resilient
  secure boot architecture with onboard recovery for risc-v based soc,''
  https://arxiv.org/pdf/2101.06300.pdf, 2020.

\bibitem{toctou:2006}
Mitre, ``Cwe-367: Time-of-check time-of-use (toctou) race condition,''
  https://cwe.mitre.org/data/definitions/367.html, 2006.

\bibitem{Hacl}
\BIBentryALTinterwordspacing
J.-K. Zinzindohou\'{e}, K.~Bhargavan, J.~Protzenko, and B.~Beurdouche, ``Hacl*:
  A verified modern cryptographic library,'' in \emph{Proceedings of the 2017
  ACM SIGSAC Conference on Computer and Communications Security}, ser. CCS
  '17.\hskip 1em plus 0.5em minus 0.4em\relax New York, NY, USA: Association
  for Computing Machinery, 2017, p. 1789–1806. [Online]. Available:
  \url{https://doi.org/10.1145/3133956.3134043}
\BIBentrySTDinterwordspacing

\end{thebibliography}

\end{document}